\documentstyle[preprint,aps,graphicx]{revtex}

\begin{document}
\draft

\title{Electron Interference Effects on the Conductance of
Doped Carbon Nanotubes}

\author{Alain Rochefort$^*$ and Phaedon Avouris$^\dagger$}

\address{$^*$ Centre de recherche en calcul appliqu\'e (CERCA), Groupe
Nanostructures et Biomat\'eriaux, 5160 Boul. D\'ecarie, bureau 400, Montr\'eal, 
(Qu\'ebec) Canada H3X 2H9.}
\address{$^\dagger$ IBM Research Division, T.J. Watson Research Center,
P.O. Box 218, Yorktown Heights, NY 10598, USA.}
\maketitle
\centerline{(\today)}

\begin{abstract} 
We investigate the effects of impurity scattering on the conductance
of metallic carbon nanotubes as a function of the relative separation
of the impurities. First we compute the conductance of a clean (6,6)
tube, and the effect of model gold contacts on this conductance.
Then, we compute the effect of introducing a single, two, and three
oxygen atom impurities. We find that the conductance of a
single-oxygen-doped (6,6) nanotube decreases by about 30 \% with
respect to that of the perfect nanotube. The presence of a second
doping atom induces strong changes of the conductance which, however,
depend very strongly on the relative position of the two oxygen
atoms. We observe regular oscillations of the conductance that repeat
over an O-O distance that corresponds to an integral number of half 
Fermi-wavelengths ($m\lambda_F/2$). These fluctuations  reflect strong
electron interference phenomena produced by electron scattering from
the oxygen defects whose contribution to the  resistance of the tube
cannot be obtained by simply summing up their individual
contributions.\\  
\end{abstract}

\section{Introduction}

Carbon nanotubes (CNTs) have very interesting electrical properties. 
Depending on their diameter and helicity it was predicted that they
can be semiconductors or metals \cite{saito,mintmire}, and this was
confirmed by scanning tunneling spectroscopy \cite{wildoer,odom}.
They can also sustain large current densities \cite{avouris}, and
their electrical properties can be modified by doping \cite{lee}.
These unique electrical characteristics coupled with their high
mechanical stability and excellent thermal conductivity make the CNTs
ideal candidates for use in nanoelectronics. Several possible
applications such as their use as channels in field-effect
transistors \cite{tans,martel}, single electron transistors
\cite{bockrath}, and diodes \cite{collins,yao} have already been
successfully demonstrated. It is therefore very important that a
detailed understanding of electrical transport and energy dissipation
in CNTs be developed.  In addition to the quantized resistance due to
the mismatch of the number of transmission channels in the tube and
the metal contacts \cite{datta}, additional sources of resistance are
provided by the formation of Schottky barriers at the contacts
\cite{leonard,xue}, and by electron scattering from adsorbed or
embedded impurity atoms and defects. A number of theoretical studies
have appeared on this last issue.  These studies have, so far,
considered only scattering by individual defects  
\cite{chico,lambin,crespi,white,ando,anantram}, or, the contributions
of a number of defects to the resistance of the tube were treated as
being additive \cite{kostyrko}. However, one important characteristic
of transport in nanotubes is  their long coherence lengths,
especially at low temperatures. This coherence allows for
interference effects involving scattering from the defect sites
present in the probed nanotube segment.\\ 

  Here we investigate two contributions to the resistance of 
nanotubes. First, we calculate the contact resistance arising from
the imperfect coupling of nanotubes with model metal electrodes. Then
we concentrate on the resistance produced by substitutional defects.
We show that the relative position of defects can have a very
important influence on the strength of scattering and on the
resulting electrical resistance.  Specifically, we calculate the
attenuation of the transmission of a metallic nanotube induced by
scattering from individual, pairs, and triplets of oxygen defect
sites as a function of their relative separation along the nanotube
axis. Oxygen atoms are used here as model defects, but are likely to
be introduced in nanotubes by  oxidative purification of the
nanotubes \cite{he,kuznetsova} or sonication.\\

\section{Computational Details}

The nanotube model used in the computations contains 948 carbon atoms
(96 {\AA} long) forming an armchair $(6,6)$ nanotube. The bond
distance between carbon atoms of the NT is fixed to that in graphite
1.42 {\AA}. This tube is bonded with its two dangling bond bearing
ends to two metal electrodes \cite{rochefort2}. Each electrode is
modeled by a layer of 22 gold atoms in a (111) crystalline
arrangement. The electrical transport properties of a system can be
described in terms of the retarded Green's function
\cite{datta,economou}.  The transmission function is computed by using
the Landauer-B\"uttiker formalism as described in detail in ref. 12,
and the effects of the semi-infinite electrodes are described by
self-energies. The Green's function ${\mathcal{G}}_{NT}$ can be
written in the form of block matrices separating explicitly the
molecular Hamiltonian:  
\begin{equation} 
{\mathcal{G}}_{NT}=\big[E{\mathcal S}_{NT}-{\mathcal
H}_{NT}-\Sigma_1-\Sigma_2 \big]^{-1}  
\end{equation}  

\noindent
where ${\mathcal S}_{NT}$ and ${\mathcal H}_{NT}$ are the overlap and
the Hamiltonian matrices, respectively, and $\Sigma_{1,2}$ are
self-energy terms that describe the effect of the electrodes. They
have the form $\tau_{i}^\dagger g_{i} \tau_{i}$ with $g_{i}$ the
Green's function for the isolated semi-infinite electrodes
\cite{dattamol,papa}, and $\tau_{i}$ is a matrix describing the
interaction between the NT and the Gold electrodes. The Hamiltonian
and overlap matrices were determined using the extended H\"uckel
method (EHM) \cite{yaehmop} for the system: Gold-CNT-Gold. It has been
shown that EHM gives results similar to those obtained on extended
NTs with more sophisticated methods \cite{rochefort1}. \\ 

The transmission function, $T(E)$, that is obtained from this Green's
function is given by \cite{datta}:

\begin{equation} 
T(E)=T_{21}=\textrm{Tr} [\Gamma_2 {\mathcal{G}}_{NT} \Gamma_1 {\mathcal{G}}_{NT}^\dagger ].  
\end{equation} 

In this formula, the matrices have the form:  

\begin{equation} 
\Gamma_{1,2}=i(\Sigma_{1,2}-\Sigma^\dagger_{1,2}).  
\end{equation} 

The summation over all conduction channels in the molecule allows the
evaluation of the conductance ($G(E_F)$) at the Fermi energy, i.e. for
zero bias between the electrodes, $G(E_F)=(2e^2T(E_F))/h$. \\

  Some of the configurations investigated in our oxygen-doping study
are shown at top of Figure 1(top). The first oxygen atom (dark atom in
Fig. 1) is located near the middle of the tube between the electrodes
while the position of the second oxygen atom is defined by the
spacing number, i.e. the number of carbon atoms that are separating
the two oxygen defect atoms along the zig-zag line. The distance
between two adjacent circular carbon planes is 1.23 {\AA}.\\

\section{Results and Discussion}

  Figure 1(bottom) shows the variation of conductance ($G(E_F)$), in
units of 2$e^2/h$, for the structurally perfect and oxygen-doped
(6,6) nanotubes. The conductance of the perfect tube connected to the
two model Au pads is calculated to be 2.3$e^2/h$, i.e. about 40\%
smaller than the expected 4$e^2/h$ value for a perfect CNT with ideal
contacts. \cite{chico,frank} This result shows that a sizable in
series resistance can be introduced by non-ideal contacts. It is
clear that contribution to the resistance of CNTs can obscure their
intrinsic resistance quantization \cite{rochefort2}. Depending on the
nature of the interface, contact resistances can have contributions
from many sources of non-ideality such as Schottky barriers or
surface roughness. The value of the computed contact resistance
($\sim 5~k\Omega$) in our configuration compares well with contact
resistances that can be inferred from experiments involving CNTs
end-bonded to metal electrodes\cite{dai}. \\

  Substitution of a carbon atom by an oxygen atom further reduces the
conductance to about 1.6$e^2/h$, a 30\% decrease. For comparison,
this reduction is similar than that produced by the introduction of a
single vacancy which leads to $G(E_F)$ = 1.6~$e^2/h$. A 50\%
reduction in $G$ was calculated using a somewhat different technique
for a vacancy in a (4,4) nanotube. \cite{chico} More recently, a
decrease of approximately 20\% of the conductance at $E_F$ was found
on a (10,10) tube containing a vacancy with a more sophisticated
technique \cite{choi}. We note that the reduction of the conductance
of an $(n,n)$ tube introduced by a weak scatterer decreases with
increasing $n$ \cite{chico,white}.\\

  Introduction of a second doping oxygen atom changes the
conductance. Most importantly, the magnitude of the change is not
constant but is strongly correlated with the relative position of the
second oxygen atom along the nanotube length. In fact, as Figure 1
shows, there is a strong oscillatory dependence of the conductance on
the separation between the two O atoms. The values of the conductance
$G(E_F)$ at the maxima are higher than the conductance of the CNT
with a single O defect, and increase gradually  as the O-O distance
increases. The separation between two successive maxima (or minima)
of $G(E_F)$ is equal to the width of three circular sections of the
nanotube, i.e. 3$\times$1.23 {\AA} = 3.69 {\AA}. The origin of the
oscillatory behavior of $G(E_F)$ becomes clear by considering the
electronic structure of the armchair carbon nanotube. Figure 2 shows
the highest occupied (HOMO) and lowest unoccupied (LUMO) molecular
orbitals of the (6,6) model tube. In an armchair nanotube these
orbitals cross the Fermi level at the $K$-point \cite{saito,rubio},
i.e. when $k = k_F = 2\pi /3a$, where  $a = \sqrt 3 R_{C-C} =$ 2.46
{\AA}. Thus, the Fermi wavelength is  $\lambda_F = 3a =$ 7.4 {\AA}
(see Fig. 2). We now see that the spacing between successive maxima
(minima) of $G(E_F)$ corresponds to half a Fermi wavelength of the
perfect nanotube. \\

 In contrast to the strong dependence of G on the O-O separation, we
find only a weak dependence on the angle between the two O atoms. A
calculation of $G$ in which the O-O separation was set at $8\pi/k_F$
(29.5 {\AA}), and where one O atom remained fixed while the second O
atom was moved around a circular carbon section gave only a weak
oscillation of $G(E_F)$ between the values of 1.87 and 1.91 $e^2/h$.
This invariance to a $C_6$ ($\pi/3$) rotation can be understood by
considering the frontier orbitals in Fig. 2 from which we see that
such a symmetry operation leaves the wave function unchanged.\\

The effect of O atom substitution on the electronic structure of the
CNT can be seen in Figure 3 which shows wave function contours of
nanotube circular sections for the perfect tube (3A) and for the tube
with two O atoms separated by 5 carbon ring sections, i.e. by 7.4
{\AA} (3B). It can be seen that the effect of oxygen substitution is
quite localized in the vicinity of the oxygen atoms. Further removed
regions (not presented) show very similar wave function contours for
both pure and doped CNTs. The main effect of O-doping is the
generation of positive charges largely localized on the adjacent C
atoms. We can then consider the two O atoms as forming a quasi-1D
potential well with a length $d$ defined by the O-O atom separation
that can scatter the Fermi level electrons. The transmission function
$T$ of such a system can then be written as \cite{merzbacher}: $T =
[1 + \textrm{A sin}^2(k'd)]^{-1}$, where A depends on the ratio of
the wavevectors of the incident wave ($k$) and of the wave inside the
potential well ($k'$). Transmission maxima will occur when $k'd =
m\pi $, and minima when $k'd = (2m-1)\pi/2$ (where $m$=1,2,3\ldots).
Thus, $G(E_F)$ will vary as a cos$^2(k'd)$ with varying O-O atom
separation $d$. The cos$^2(k'd)$ envelope is shown by the dot-dashed
line in Figure 1 (we have assumed that $k = k' = k_F$). \\

It is clear that the simple model involving electron interference of Fermi
level electrons scattered by the potential well formed by the two O atoms can
qualitatively explain the main features of Figure 1.  However, deviations
from this simple picture are also  evident, and are most significant when the
two O atoms are close to each other.  First, in the 1-D potential well model,
impurities separated by a distance of $m\lambda_F/2$ should become
transparent to the incident electron waves, i.e. $T =$ 1 \cite{merzbacher},
but although the transmission is indeed maximized at these separations, it
never reaches unity.  By comparing the effect of one O atom with the effects
of two O atoms on $G$, and from the orbital contours of Fig.  3, we can
conclude that O doping affects the conductance in two ways.  Introduction of
the first O atom into the CNT introduces a change in the local electronic
structure which decreases the conductance of the tube. The introduction of
the of the second O atom which reduces the symmetry to C1 leads to complex
changes in the conductance of the tube. These changes depend on the relative
distance between the two O atoms. The resulting contribution to the
resistance due to the change in electronic structure is large, particularly
when the two O atoms are close together, indicating a cooperative distortion
of the electronic structure. At larger O-O separations, backscattering from
the well becomes more important and $G$ shows a clear O-O separation
dependence as a result of interference between incident and back-reflected
electrons in the well.\\

The interference effects observed with the two oxygen atom models can
be generalized to CNTs doped with larger numbers of dopant atoms. The
three oxygen atom case is particularly interesting. Some results are
shown in Fig. 1 (open circles) where we have fixed the spacing
between the two first oxygen atoms (O(1) and O(2)) to correspond to a
constructive interference resonance (4$\pi/k_F$), and varied the
position of the third O atom (O(3)). Again, oscillations in $G(E_F)$
are observed with maxima at O(2)-O(3) separations equal to
$m\pi/k_F$, and minima when this distance is $(2m-1)\pi/2k_F$. The
values of $G(E_F)$ computed for the three O atoms case are within the
range found for the two O atoms case (Fig. 1). However, the same
resonance is not observed when the outer two O atoms, O(1) and O(3)
are kept at a distance corresponding to constructive interference,
e.g. at 8$\pi/k_F$, and O(2) is placed between them at a distance
corresponding to destructive interference, e.g. at 7$\pi/2k_F$ from
O(1). The O(2) atom causes a strong damping of the resonance leading
to a $G(E_F)$ of only 0.98$e^2/h$.\\

Next we consider the relation between $G(E_F)$ and the density of
states at the Fermi energy, $DOS(E_F)$. The Drude conductivity of
solids is proportional to the density of states, and a similar 
correlation was found by first-principles calculations on molecular
wires \cite{lang}. As Fig. 4 shows, introduction of oxygen atoms in
the armchair nanotube increases the $DOS(E_F)$. However, Fig. 4 also
shows that there is an anticorrelation between $G(E_F)$ and
DOS$(E_F)$, the latter exhibiting approximately a sin$^2({k_F}d)$
dependence while the former shows a cos$^2({k_F}d)$ dependence. In
Fig. 5 we also show the computed  local density of states (LDOS) for
nanotubes containing no oxygen atom (A), one oxygen atom (B), and two
oxygen atoms located at distances leading to constructive (C) and
destructive interference (D), respectively. These LDOS values
represent the sum of the contributions to the density of states of
the first three C atoms adjacent  to the O impurity and of the
impurity itself (full line). The contribution of the O atom alone is
represented by the dashed-dotted line.  The clean nanotube LDOS (5A)
shows the first two van Hove singularities on either side of
$E_F$. Upon introduction of the first O atom a new quasi-bound state
is formed centered at about 0.3 eV above $E_F$ with a tail that
extends to $E_F$ (5B). This state is quite similar to that produced
by the introduction of nitrogen \cite{choi}. Position dependent
modifications of the electronic structure with respect to Fermi
energy are observed upon introduction of the second O atom (5C and
5D).\\

The behavior of $DOS(E_F)$ in Fig. 4 can be understood by considering
the changes in bonding produced by the substitution of a C atom by an
O atom. This substitution generates non-bonding states whose center 
of gravity is near $E_F$. Thus, although there is an increase in
$DOS(E_F)$, the conductance does not increase because of the
localized nature of these O-induced states. This leads to an
anti-correlation between $DOS(E_F)$ and $G(E_F)$. Turning to the
behavior of the LDOS (Fig. 5), we note that there is a minimum at the
O site when the O-O separation corresponds to a resonance
(7$\pi/{k_F}$). This is likely due to the formation of the nodal
front of the standing wave between the O(1) and O(2) atoms (Fig. 5C).
The LDOS in Fig. 5D where the O-O separation (15$\pi/{2k_F}$) leads
to destructive interference, shows no such minimum. \\

\section{Conclusions}

In conclusion, we have shown that interference effects involving
scattering from pairs of defects in carbon nanotubes and
defect-defect interactions can have a strong influence on the
electrical resistance of the tubes. Due to the long coherence lengths
in carbon nanotubes, the net contribution of a number of scatterers
cannot be determined merely by the sum up of their individual
contributions.\\

\section{Acknowledgment}

We thank Fran\c{c}ois L\'eonard and Massimiliano Di Ventra for their
helpful comments on the manuscript.

%
%
\clearpage
\newpage
\begin{figure}[p]
\vspace{1.0cm}
\caption{Top: Schematic showing the different positions of two oxygen
atom dopants in the (6,6) carbon nanotube model. The position of the
first atom (dark) is fixed, while the possible positions of the
second O-atom are indicated by the numbers. Bottom: Computed
conductance (in units of 2$e^2/h$) of a (6,6) nanotube under
diffferent conditions. The dotted line indicates the conductance of
the clean nanotube that  includes a series resistance due to the
imperfect contacts with gold pads (see text). The second dotted line
shows the conductance after the incorporation of a single oxygen
atom.  The solid circles give the conductance of the tube after the
incorporation of a second oxygen atom as a function of the separation
between the two O-atoms. The empty circles give the conductance of
the tube doped with three oxygen atoms when the distance between the
first and the second O-atoms is fixed at 4$\pi/k_F$ (14.8 {\AA}) and
the position of the third is varied.}  

\vspace{1.0cm}
\caption{Representation of: (A) the highest occupied molecular orbital
(HOMO), and (B) the lowest unoccupied (LUMO) orbital of an undoped (6,6)
armchair nanotube model.}

\vspace{1.0cm} 
\caption{Comparison of the orbital contours of the highest occupied orbital
(HOMO) of (A): a clean (6,6) tube, and (B): a tube doped by two oxygen atoms 
(dark circles). The contours are generated in a plane perpendicular to the 
nanotube axis.}

\vspace{1.0cm}
\caption{Variation of the total density of states at the Fermi level
$DOS(E_F)$ of oxygen-doped nanotubes. The dot-dashed line shows a 
sin$^2({k_F}d)$ envelop}

\vspace{1.0cm}
\caption{Local density of states (LDOS) in the vicinity of the
oxygen impurities in a (6,6) carbon nanotube containing no
oxygen atom (A), one oxygen atom (B),
two O atoms located at a resonance position (C), and at an
antiresonance position (D). Full lines represent the DOS
contributions from the impurity itself and the first three neighboring
carbon atoms, while dashed-dotted lines give the DOS of the
impurity only. The O-O separation is given in panels C and D.} 

\end{figure}
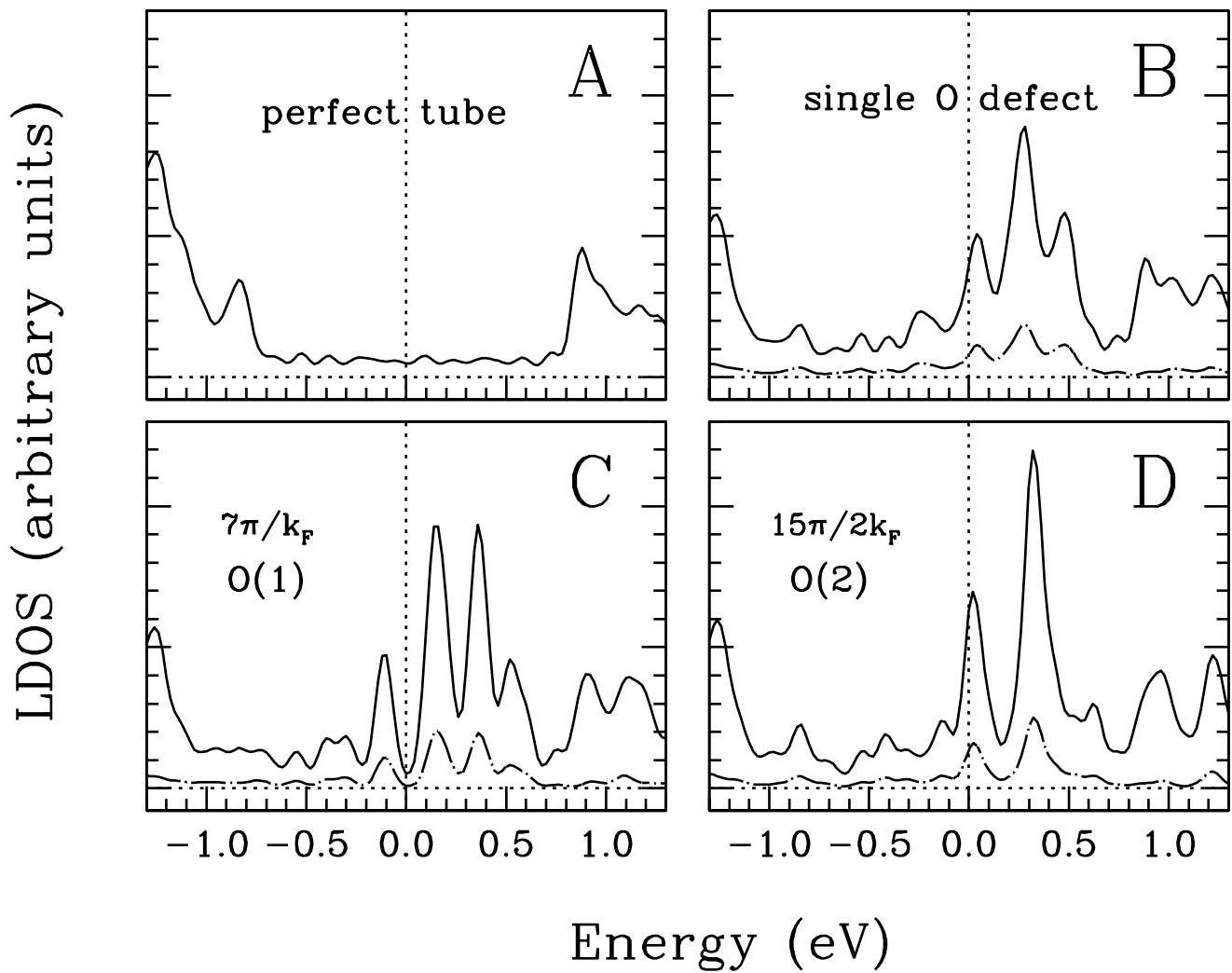
\end{document}